# Phenomenological model for the direct and inverse Edelstein effects

Hironari Isshiki[1], Prasanta Mudli[1], Junyeon Kim[2], Kouta Kondou[1,2], and YoshiChika Otani[1,2,3]

[1] Institute for Solid State Physics, University of Tokyo, Kashiwa, Chiba 277-8581, Japan

[2] Center for Emergent Matter Science, RIKEN, Wako, Saitama 351-0198, Japan

[3] Trans-scale Quantum Science Institute, University of Tokyo, Bunkyo-ku, Tokyo 113-0033, Japan

**(Abstract)**

We have developed a phenomenological model that connects the direct and the inverse Edelstein effects. Our model implies a trade-off relation between the conversion coefficients for the direct and inverse effects. Thus, a large conversion coefficient for the inverse effect does not necessarily bring a large conversion coefficient for the direct effect. Instead of these coefficients, we propose a figure of merit of Edelstein effects which consists of two factors; one of them represents the magnitude of the spin-orbit coupling, and the other represents the strength of the hybridization between bulk and interface states. Both of them are quit important for the efficient conversion through Edelstein effects. To test our model, we measured the inverse and direct Edelstein effects at the $Bi_2O_3$/Cu interface using spin absorption method with a non-local spin valve structure and calculated the conversion coefficients. The effective spin Hall angle reaches ~0.09 in this system. This relatively large value is attributable to not only the large spin-orbit coupling but also the strong hybridization between the interface and bulk states at the $Bi_2O_3$/Cu interface.

**(Main text)。**

The spin-charge current interconversion based on the Edelstein effect (EE) in interface [1,2] has attracted much interest recently due to its qualitatively different conversion mechanism compared to the spin Hall effect (bulk effect) [3]. These conversions are attributable to the spin-momentum locking at the interface, the orthogonal coupling between spins and momenta of electrons caused by the Rashba effect [4], or the topological surface states [5]. Unfortunately, the characterization of the conversion efficiency has always relied on either inverse or direct Edelstin effect. Still, most of the reports ignore to evaluate both of them in the same device. Some attempts have already been made to study both conversions simultaneously in the same device employing static and dynamical spin injection measurements. However, we here focus on static spin injection analysis, such as the Edelstein Magnetoresistance [6,7]. An analytical model was developed by considering the spin relaxation at the interface and the adjacent bulk [7] to understand the Edelstein magnetoresistance. However, some problems remain; the model cannot yield the conversion coefficient for the direct effect. Besides, the resistivity and thickness of the interface were arbitrary. Furthermore, there is a contradiction in the analysis *i.e.* the model assumes a momentum relaxation constant with variable thickness of the metallic layer, while the Rashba parameter changes with the thickness-dependent interface crystallinity. In the present work, to solve the above problems, we formulate a new phenomenological model connecting the direct and inverse conversions and propose figures of merit for them. Our model is beneficial for evaluating the efficiency of both Edelstein effects. We test our model in the analysis of the modified non-local spin valve measurements, which enables us to study both conversions on the same devices so that we can obtain both conversion coefficients. We chose a $Bi_2O_3$/Cu bilayer [7–11] as a test interface to determine the preferable conditions for efficient conversion.

We consider the system consists of a bulk state and an interface state as shown in Fig. 1(a). The direct and inverse Edelstein effects occur due to the transmission of electrons between these two states. A spin accumulation in the interface relaxes by two independent processes; (1) the spin relaxation accompanies the momentum scattering in the interface state, (2) the spin relaxation due to the leakage of the spin from the interface to the bulk state. We define two relaxation times $\tau_\mathrm{p}$ and $\tau_\mathrm{t}$ which characterize these two relaxations, *i.e.* the spin relaxation time in the interface state and the spin transmittion time across the interface, respectively, as drawn in Fig. 1(b)).

In the direct Edelstein effect (DEE), applied 2D charge current ($j_\mathrm{c_DEE}{}^\mathrm{2D}$ [A/m]) produces a spin accumulation at the interface, and a part of the accumulated spins escapes into the bulk as diffusive 3D spin current ($j_\mathrm{s_DEE}{}^\mathrm{3D}$ [A/m$^2$]) with spin polarization direction orthogonal to the $j_\mathrm{c_DEE}{}^\mathrm{2D}$. The charge-to-spin current conversion coefficient is thus defined as $q \equiv j_\mathrm{s_DEE}{}^\mathrm{3D}/j_\mathrm{c_DEE}{}^\mathrm{2D}$ [m$^{-1}$] [12]. However, especially for DEE, which occurs through the Rashba effect, the physical parameter that characterizes $q$ has not been clarified yet. We here deduce the phenomenological expression for $q$. The total spin accumulation produced by $j_\mathrm{c_DEE}{}^\mathrm{2D}$ at an interface with an $\alpha_\mathrm{R}$ is approximated as [13]

$$\langle\delta S\rangle \sim \frac{m_\mathrm{2D}\alpha_\mathrm{R}}{ge\hbar E_\mathrm{F}} j_\mathrm{c_DEE}{}^\mathrm{2D}, \quad (1)$$

where $g$ is Landé g-factor, $e$ the elementary charge, $m_\mathrm{2D}$ the effective mass of the electron in the interface state, and $E_\mathrm{F}$ the Fermi energy. Note that Eq. (1) is valid for “high-density regime” where the Fermi energy is much larger than the spin-orbit coupling energy $E_\mathrm{F} \gg \alpha_\mathrm{R}{}^2 m_\mathrm{2D}/2\hbar^2$ [14]. When the Fermi energy is comparable to the energy of the spin-orbit coupling, $E_\mathrm{F} \leq \alpha_\mathrm{R}{}^2 m_\mathrm{2D}/2\hbar^2$, low-density regime [14], the perturbative treatment of the spin orbit coupling would fail, and therefore Eq. (1) are not valid anymore [14,15]. In this study, we only consider the high-density regime. For generating $j_\mathrm{s_DEE}{}^\mathrm{3D}$ from $\langle\delta S\rangle$ at the interface, there must be a part of accumulated spins escaping from the interface to the bulk, as shown in Fig. 1(a). Therefore, $j_\mathrm{s_DEE}{}^\mathrm{3D}$ should be proportional to the total spin accumulation and inversely proportional to a spin transmission time across the interface $\tau_\mathrm{t}$;

$$j_\mathrm{s_DEE}{}^\mathrm{3D}/e = \langle\delta S\rangle/\tau_\mathrm{t}. \quad (2)$$

From Eqs. (1) and (2) with $E_{\mathrm{F}} \sim \frac{1}{2}mv_{\mathrm{F}}^{2}$ ($v_{\mathrm{F}} = \hbar k_{F}/m$, $k_{F}$ is the Fermi wavevector at the interface state appropriate in the absence of the spin-orbit coupling [16]), we find that the conversion coefficient $q$ can be written as,

$$q \equiv \frac{j_{\mathrm{s_DEE}}{}^{\mathrm{3D}}}{j_{\mathrm{c_DEE}}{}^{\mathrm{2D}}} \sim \frac{\alpha_{\mathrm{R}}}{v_{\mathrm{F}}{}^{2}\hbar\tau_{\mathrm{t}}} \ [\mathrm{m}^{-1}]. \ (3)$$

In the inverse Edelstein effect (IEE), the injected 3D spin current ($j_{\mathrm{s_IEE}}{}^{\mathrm{3D}}$ [A/m$^2$]) into the interface produces a spin accumulation and results in the 2D charge current ($j_{\mathrm{c_IEE}}{}^{\mathrm{2D}}$ [A/m]). The balance between the injection of spin current and the relaxation of the spin in the interface can be expressed as [2],

$$j_{\mathrm{s_IEE}}{}^{\mathrm{3D}}/e = \langle\delta S\rangle/\tau_{\mathrm{p}}. \ (4)$$

The spin current defined here is equal to that we can estimate experimentally. The 2D charge current density associated with the spin accumulation is

$$j_{\mathrm{c_IEE}}{}^{\mathrm{2D}} = \frac{\alpha_{\mathrm{R}}e}{\hbar}\langle\delta S\rangle\left(1 - \frac{\frac{1}{\tau_{\mathrm{t}}}}{\frac{1}{\tau_{\mathrm{p}}}+\frac{1}{\tau_{\mathrm{t}}}}\right). \ (5)$$

Here, $\hbar$, $\alpha_{\mathrm{R}}$ and $e$ are the Dirac constant, the Rashba parameter of the interface and elementary charge, respectively. The factor in the parenthesis in the right hand side represents the decrease of the charge current due to the additional spin leakage to the bulk state. From Eq. (4) and (5), the conversion coefficient for IEE known as Edelstein length is obtained as,

$$\lambda \equiv \frac{j_{\mathrm{c_IEE}}{}^{\mathrm{2D}}}{j_{\mathrm{s_IEE}}{}^{\mathrm{3D}}} \sim \frac{\alpha_{\mathrm{R}}\tau_{\mathrm{IEE}}}{\hbar} \ [\mathrm{m}], \ (6)$$

$$\frac{1}{\tau_{\mathrm{IEE}}} = \frac{1}{\tau_{\mathrm{p}}} + \frac{1}{\tau_{\mathrm{t}}}. \ (7)$$

In the interface of topological insulator and metal, the similar expression for $\tau_{\mathrm{IEE}}$ has been deduced by Dey *et al.* by solving a Boltzmann equation [17]. The relaxation time $\tau_{\mathrm{IEE}}$ can be interpreted as the modified momentum relaxation time due to the additional spin relaxation in the bulk state. On the other hand, in the model developed by Sanz-Fernandez *et al.*, Edelstein length was described by the interfacial spin-charge conductivity and the spin-loss conductance.

We should note that the magnitudes of $\lambda$ and $q$ are governed by different relaxation times, $\tau_{\mathrm{IEE}}$ and $\tau_{\mathrm{t}}$, respectively. Besides, $\tau_{\mathrm{IEE}}$ and $\tau_{\mathrm{t}}$ correlate to each other. We find a trade-off relation between $\lambda$ and $q$, represented. The interface with a large $\lambda$ brings a small $q$. For example, at the surfaces of single-crystal metal such as Cu(111) [18,19], $\tau_{\mathrm{IEE}}$, $\tau_{\mathrm{p}}$ and $\tau_{\mathrm{t}}$ would be large because of the perfect crystallinity and the well separated surface and bulk states at the Fermi level. In the spin-charge current interconversions at these surfaces, the large $\tau_{\mathrm{IEE}}$ and $\tau_{\mathrm{t}}$ would contribute to large $\lambda$ and small $q$, respectively. The similar situation is expected at the epitaxially grown interfaces. Vaz *et al.* claimed that the epitaxial $AlO_x$/STO interface gives an optimum value of $\lambda$ as a result of much larger $\tau_{\mathrm{t}}$ than $\tau_{\mathrm{p}}$ [20]. However, in this system, $q$ would be small due to the large $\tau_{\mathrm{t}}$. One major issue of the Edelstein effects is an efficient spin current generation to switch the magnetization of the adjacent ferromagnetic layer via the spin-transfer torque [21]. The interfaces discussed above may not be efficient spin generators due to the large $\tau_t$. Vaz *et al.* also discussed the application potential of this system for the spin current detector due to an expected high output voltage in the IEE. However, a large $\tau_t$ means the small pass for the spin current, which would hamper the injection of the spin current into the interface and limit the output voltage. On the other hand, the interfaces with small $\tau_t$ may show large $q$ and small $\lambda$. Thus, only discussing $\lambda$ or $q$ is not sufficient to understand the interconversion via EE.

We should take into account both $\lambda$ and $q$ to examine the efficiency of the Edelstein effects. We here consider the product of $\lambda$ and $q$;

$$\lambda \cdot q = \left(\frac{\alpha_R{}^2}{v_F{}^2\hbar^2}\right)\frac{\tau_{IEE}}{\tau_t} = \left(\frac{\alpha_R{}^2}{v_F{}^2\hbar^2}\right)\frac{\tau_p}{\tau_p + \tau_t} \sim \theta_{eff}{}^2. \quad (8)$$

The factor in the parenthesis represents the strength of the spin-orbit coupling normalized by the Fermi energy, which should be smaller than one in the high-density regime ($E_F \gg \alpha_R{}^2 m_{2D}/2\hbar^2$). The ratio $\tau_p/(\tau_p + \tau_t)$ is also smaller than one. Thus, the product must be no more than one and can be an indicator of the magnitude of the spin-charge current interconversion at the interface. This product seems to be nearly equivalent to a squared effective spin Hall angle $\theta_{eff}{}^2$.

$$\theta_{eff} \sim \frac{\lambda}{v_F\sqrt{\tau_{IEE}\tau_p}} \sim q v_F\sqrt{\tau_{IEE}\tau_p}. \quad (9)$$

Our model implies that the ratio $\tau_p/(\tau_p + \tau_t)$ is as important as the strength of the spin-orbit coupling for the EE. To create the interface with the large effective spin Hall angle, small $\tau_t$ and large $\tau_p$ are preferable. This ratio represents the strength of the hybridization between the interface and the bulk states. With a nearly completely isolated interface state ($\tau_t \sim \infty$), we can not inject or extract any spin current across the interface regardless of the strength of the spin-orbit coupling. Thus, such interfaces would be useless for spin-charge current intetrconversion and $\theta_{eff}$ would be zero.

When we employ the surface state of topological insulators or Rashba effect with the low-density regime ($E_F \leq m_{2D}\alpha_R{}^2/2\hbar^2$), the parenthesis factor in Eq. (8) would be larger than one. Therefore, a more efficient conversion could occur at these interfaces. The Edelstein effects in the surface states of topological insulators yield coefficients $\lambda$ and $q$ as,

$$q = 1/v_F\tau_t \ [\mathrm{m}^{-1}], \quad (10)$$

$$\lambda = v_F\tau_{IEE} \ [\mathrm{m}], \quad (11)$$

which are obtained by the substituting $\alpha_R = v_F\hbar$ in Eq. (3) and (6), respectively [24]. Apart from that, there is an approach to increase $\theta_{eff}$ by decreasing $\tau_t$; enhancing the spin transmission across the interface and bulk states by using exchange coupling, *i.e.* putting ferromagnets directly on top of the interface state. Substantial spin Hall angle has reportedly appeared in these systems [22,23] .

We have tested our model by using an oxide/metal interface in which the strong orbital hybridization between Cu and Bi has been reported [11]. We measured DEE and IEE at $Bi_2O_3$/Cu interface using non-local spin absorption method [25] with a $Bi_2O_3$(10 nm)/Cu(12.9 nm – 23.1 nm) middle wire inserted in between the two NiFe wires of a non-local spin valve device. The configurations for IEE and DEE measurements are illustrated in Fig. 2 (a) and (b), respectively. The details about the device fabrication and the geometrical parameters are described in the supplemental information. The measurements were carried out at 10 K with lock-in amplifier. In the IEE measurement, charge current $I_{C_IEE}$ (= 500 μA) was applied between NiFe and Cu wires, the created spin current diffuses along Cu wire and injected into the $Bi_2O_3$/Cu interface in the middle wire. The spin current was converted to the charge current via IEE and results in non-local voltage $V_{IEE}$ at both ends of the middle wire. In the DEE measurement, charge current $I_{C_DEE}$ (= 500 μA) was applied to the middle wire and the generated spin current by DEE at the $Bi_2O_3$/Cu interface propagates diffusively through the Cu wire (bridge wire) and is absorbed into the NiFe wires. The non-local voltage $V_{DEE}$ is detected between NiFe wire and Cu wire in the non-equilibrium steady state. The IEE (DEE) resistance was defined as $R_{IEE} = V_{IEE}/I_{C_IEE}$ ($R_{DEE} = V_{DEE}/I_{C_DEE}$). Recorded $R_{IEE}$ and $R_{DEE}$ with different $t_{Cu}$ as the function of magnetic field $H$ are plotted in Fig. 2 (c). We found that the amplitude of Edelstein resistances ($\Delta R_{IEE}$ and $\Delta R_{DEE}$) indicated by bi-direction arrow are exactly same for DEE and IEE in each device, for different $t_{Cu}$, which is consistent with the result of spin absorption measurements spin Hall materials [26]. In our experiment, $\Delta R_{IEE}$ and $\Delta R_{DEE}$ reach the greatest value with 0.080 mΩ when the copper thickness $t_{Cu}$ is 18.9 nm sample.

Before considering DEE and IEE, we would like to discuss the effective conversion efficiency of $Bi_2O_3$/Cu middle wire, in which the middle wire is assumed to be a spin Hall material with effective inverse and direct spin Hall angles $\theta^*_{\mathrm{ISHE}}$ and $\theta^*_{\mathrm{DSHE}}$, respectively. We used the equations described in Ref. [27] to calculate $\theta^*_{\mathrm{ISHE}}$, which were based on 1D spin diffusion model. The shunting factors $x$ for our $Bi_2O_3$/Cu wires were calculated numerically with *spinflow3D*. We also calculated $\theta^*_{\mathrm{DSHE}}$ form the direct measurement which has not been attempted so far (see supporting materials for the detail of the calculations). The absolute values of $\theta^*_{\mathrm{DSHE}}$ and $\theta^*_{\mathrm{ISHE}}$ as the function of $t_{Cu}$ are shown in Fig. 3(a) (Both of $\theta^*_{\mathrm{DSHE}}$ and $\theta^*_{\mathrm{ISHE}}$ are negative). We found that $\theta^*_{\mathrm{DSHE}} = \theta^*_{\mathrm{ISHE}}$ for all of our samples, which is consistent with the Onsagar's reciprocity reported in Ref. [26] and natural consequence from the experimental results: $\Delta R_{DEE} = \Delta R_{IEE}$. Note that $\theta^*_{\mathrm{DSHE}}$ and $\theta^*_{\mathrm{ISHE}}$ were calculated independently from the direct and inverse measurements, respectively. Thus, the result $\theta^*_{\mathrm{DSHE}} = \theta^*_{\mathrm{ISHE}}$ verifies our calculations of $\theta^*_{\mathrm{DSHE}}$. The maximum value of the *effective* spin Hall angle is ~ 0.09 that is relatively large and thus, $Bi_2O_3$/Cu interface has great potential to be used in spin current generation and detection.

We calculated $\lambda$, $q$, $\tau_{\mathrm{IEE}}$, $\tau_{\mathrm{t}}$ and $\tau_{\mathrm{p}}$ at our $Bi_2O_3$/Cu interface. The calculation of $q$ is more difficult than that of $\lambda$, because we have to know ${j_{\mathrm{c_DEE}}}^{\mathrm{2D}}$ in Eq. (3) (for the calculation of $\lambda$, we only need $V_{\mathrm{IEE}}$, but we don't have to know $j_{\mathrm{c_IEE}}{}^{\mathrm{2D}}$: we can assume that all of the spin current was generated at the interface). We estimate $j_{\mathrm{c_DEE}}{}^{\mathrm{2D}}$ by considering the parallel circuit of the 2D interface and the bulk Cu;

$$j_{\mathrm{c_DEE}}{}^{\mathrm{2D}} = \frac{x\rho_{\mathrm{M}}}{\rho_{\mathrm{2D}}} \frac{I_{\mathrm{C_DEE}}}{w_{\mathrm{M}} t_{\mathrm{M}}}. \quad (12)$$

$$\rho_{\mathrm{2D}} = \frac{m_{\mathrm{2D}}}{DOS_{\mathrm{2D}} E_F e^2 \tau_{\mathrm{IEE}}} \quad . (13)$$

Where $\rho_{\mathrm{2D}}$ $[\Omega]$ is the 2D resistivity of the interface, $DOS_{\mathrm{2D}} = m_{\mathrm{2D}}/\pi\hbar^2$ is the density of state of the interface state. The Fermi level of Cu ($E_F$ = 7 eV) was adopted in Eq. (13). As we have mentioned, in the previous analysis, there were arbitrariness in the 3D resistivity and the thickness of the interface. The values of $\sqrt{\lambda q}$ shown in Fig. 3(a) are similar in the order of the effective spin hall angles. This result supports the validity of our model. To obtain $\tau_{\mathrm{IEE}}$, $\tau_{\mathrm{t}}$ and $\tau_{\mathrm{p}}$, we employed the parameters reported in Bi/Cu(111) interface [28]; $\alpha_{\mathrm{R}}$ = 0.82 eVÅ and $v_{\mathrm{F}}$ = $5.21\times10^5$ m/s. We are assuming constant $\alpha_{\mathrm{R}}$ against the thickness of Cu, because the spin-orbit coupling is not sensitive against the thickness when it is larger than 13 nm, which is consistent to our previous report of $Bi_2O_3$/Cu [7,8]. See supporting materials for the detail of the calculations.

The absolute values of $\lambda$, $q$ and the relaxation times as the function of the thickness of Cu $t_{\mathrm{Cu}}$ are shown in Fig. 3(b) and (c) ($\lambda$ and $q$ are negative values). The value of $\lambda$ increases along with an increase of the thickness of Cu from 12.9 nm to 18.9 nm. This trend of $\lambda$ reflects the improvement of crystallinity of the interface with increase of $t_{\mathrm{Cu}}$, as seen in the thickness dependence of $\tau_{\mathrm{p}}$. The value of $\tau_{\mathrm{p}}$ is proportional to the momentum relaxation time of the Cu $\tau_{\mathrm{Cu}}$ ( $= m/ne^2 \rho_{\mathrm{Cu}}$) when the thickness is small, but reaches the maximum value at ~ 10 fs with $t_{\mathrm{Cu}}$ = 18.9 nm where the effect of crystallinity improvement is saturated. The maximum value of $\lambda$ (= 0.97 nm) is comparable to that reported in $Bi_2O_3$/Cu interface with spin pumping measurement [8]. On the other hand, the value of $q$ is almost constant against the change of the $t_{\mathrm{Cu}}$. This implies that the value of $\tau_{\mathrm{t}}$ (~ 30 fs) is not very sensitive to the crystallinity among the series of our samples. However, as we have mentioned above, the value of $\tau_{\mathrm{t}}$ would be several orders of magnitude larger in the epitaxially grown interfaces. According to our model, the effective spin Hall angle consists of two factors; the spin-orbital coupling factor $\alpha_{\mathrm{R}}/v_{\mathrm{F}}\hbar$ and the hybridization factor $\sqrt{\tau_{\mathrm{p}}/(\tau_{\mathrm{p}} + \tau_{\mathrm{t}})}$, which are estimated to be ~ 0.2 and ~ 0.5, respectively. Thus, hybridization factor is larger than the spin-orbital coupling factor. The large hybridization factor at Cu/$Bi_2O_3$ contributes a lot to the relatively large effective spin Hall angle.

In summary we found that the conversion coefficient of IEE is characterized by the modified momentum scattering time at the interface $\tau_{\mathrm{IEE}}$, while the coefficient of DEE is characterized by the spin transmission time across the interface $\tau_{\mathrm{t}}$. These two relaxation times are correlated by Eq. (5), which implies the trade-off relation between the conversion coefficients. Thus, a large conversion coefficients of the inverse effect does not necessarily mean the large conversion coefficients of the direct effect. We propose the product of these conversion coefficients as the figure of merit of EEs, which consists of spin-orbital coupling factor and the hybridization factor. A large ratio of $\tau_{\mathrm{p}}/(\tau_{\mathrm{p}} + \tau_{\mathrm{t}})$ is a crucial factor for the high efficiency for EE.

**Acknowledgement**

This work was supported by Grant-in-Aid for Scientific Research on Innovative Area, “Nano Spin Conversion Science” (Grant No. 26103002) and Grant-in-Aid for Young Scientists (B) (Grant No. 19K15431).

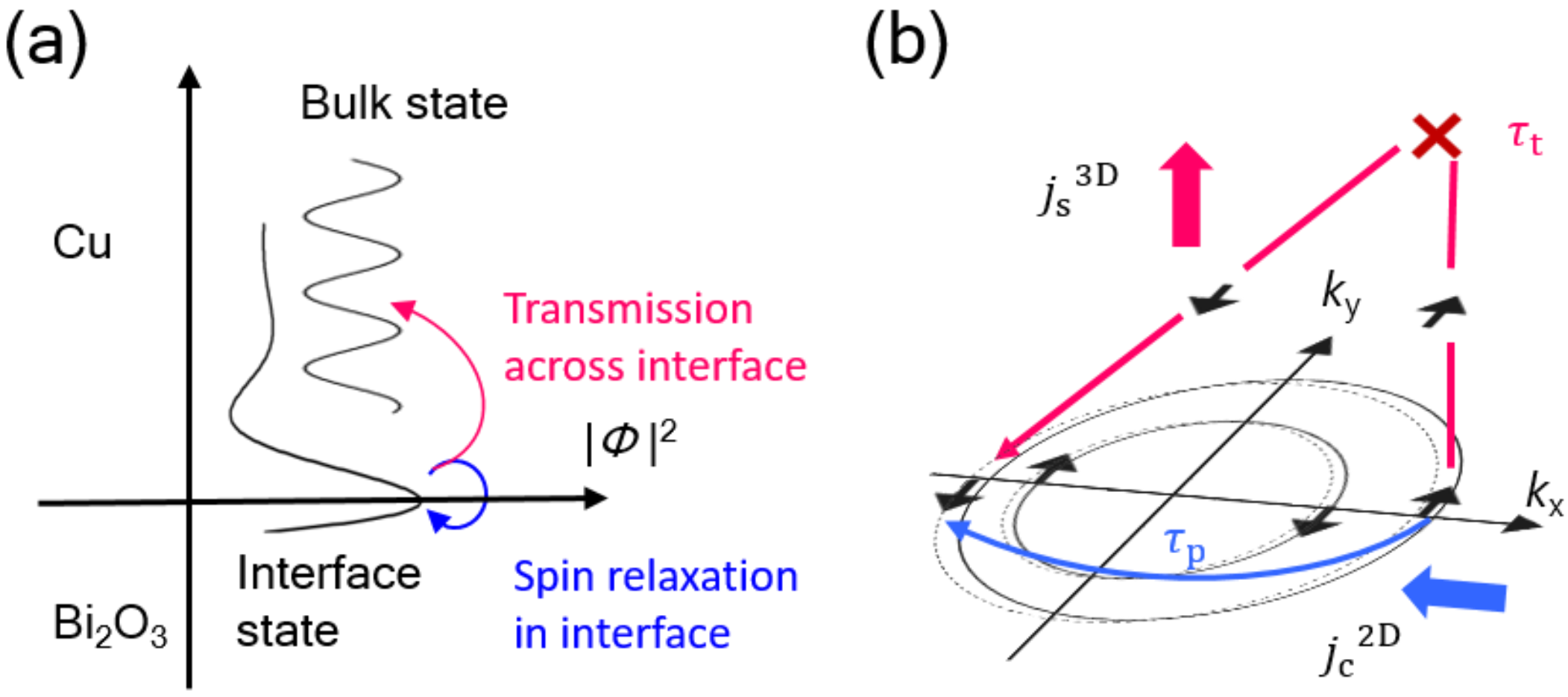

Figure 1: The concept of Edelstein effect. (a) The spin transmission across the interface and the momentum relaxation at the interface state. (b) Fermi contour of the Rashba-splitted band and illustration of direct Edelstein effect. A charge current along *x* direction create *y*-polarized spin accumulation. Blue allow indicates a momentum scattering at the interface. Magenta allows indicate the spin transmission across the interface. Red cross represents spin-flip in the ferromagnetic materials.

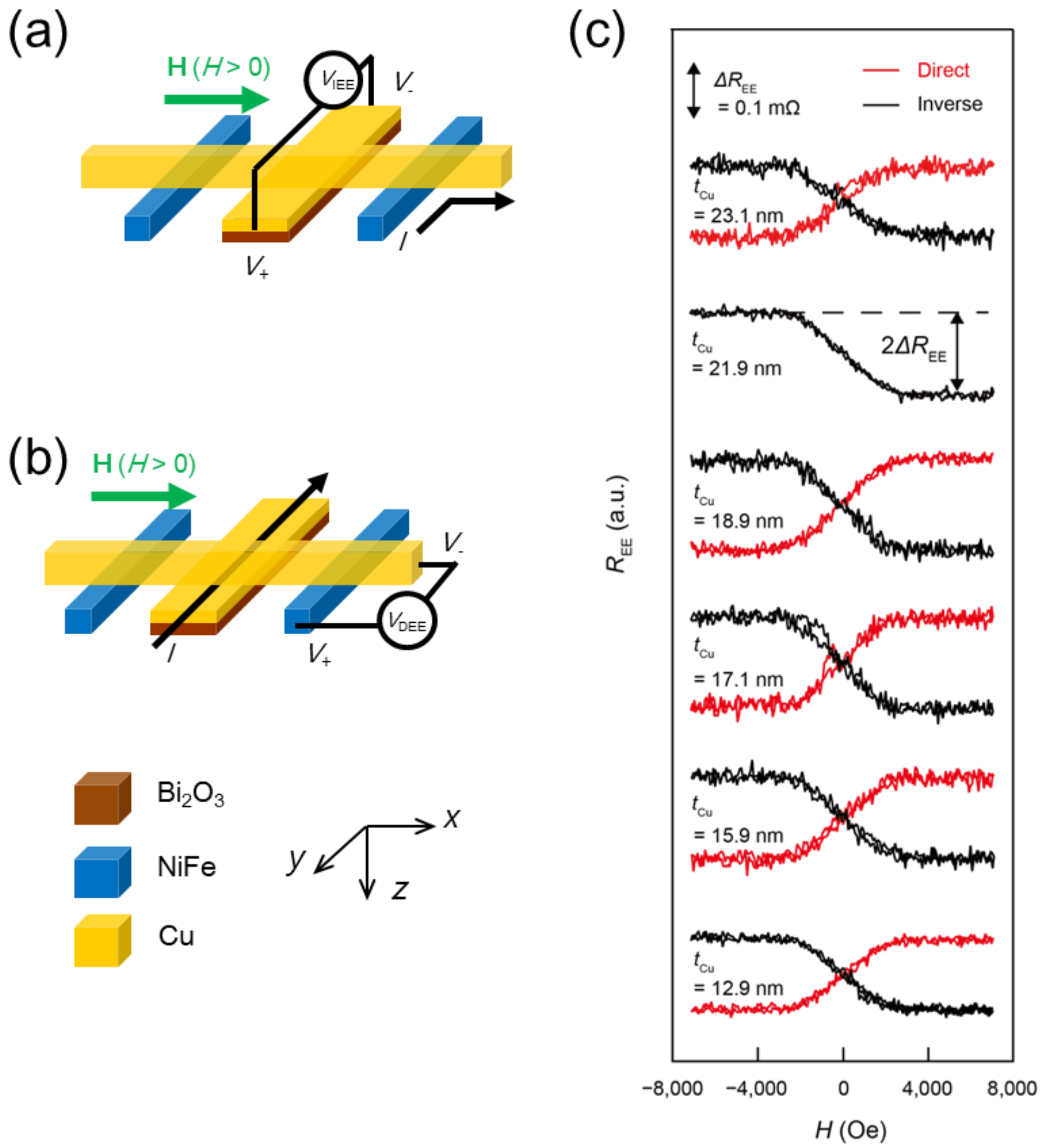

Figure 2: The measurement configuration for (a) inverse and (b) direct Edelstein effects with spin absorption method. (c) The Inverse (red lines) and direct (black lines) Edelstein resistances as the function of magnetic field at $Bi_2O_3$/Cu interface with four different Cu thickness of $Bi_2O_3$/Cu wire (The DEE measurement with $t_{Cu}$ = 21.9 nm was not done).

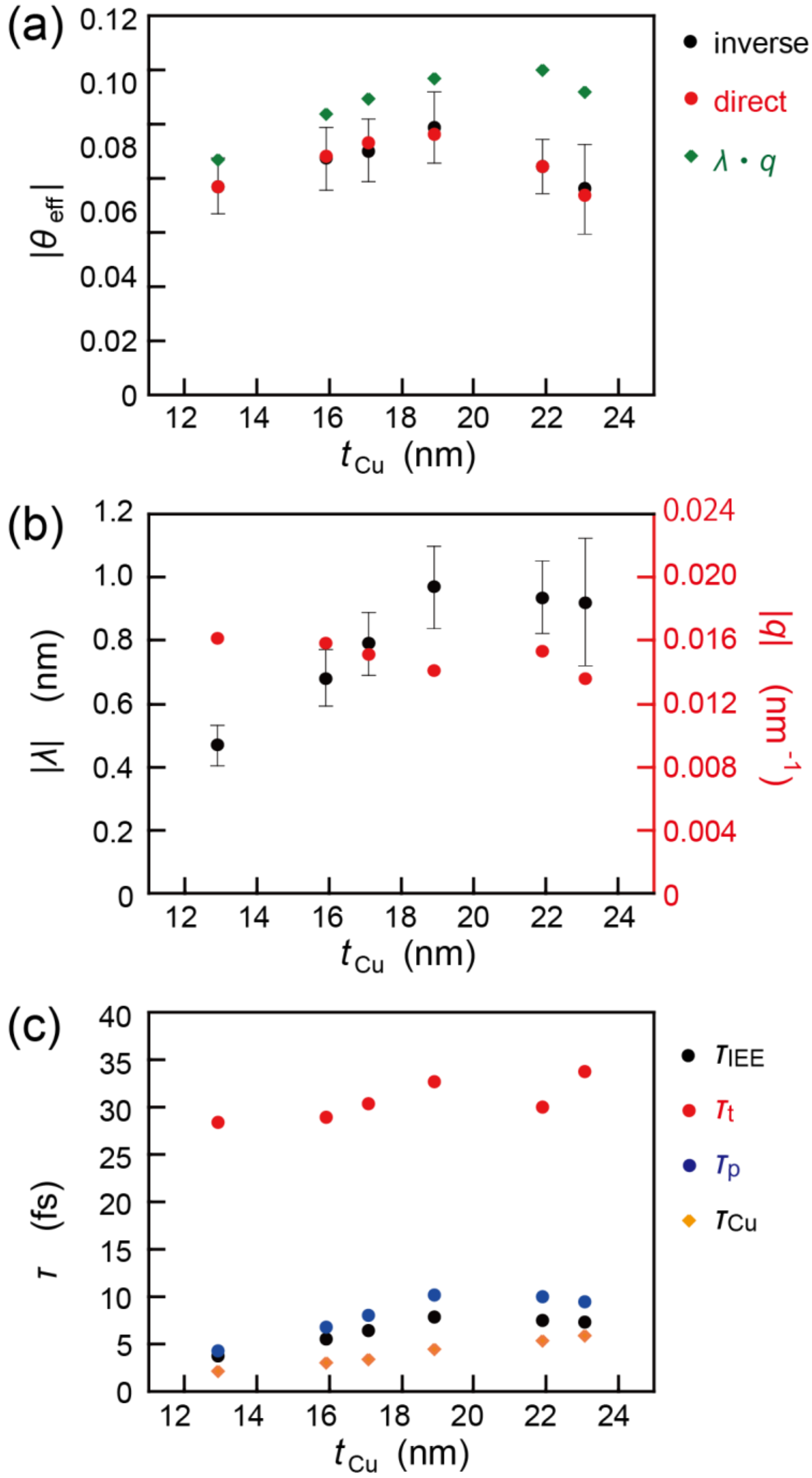

Figure 3: (a) The effective spin Hall angle, (b) the conversion coefficients of the inverse and direct Edelstein effect and (c) the relaxation times of $Bi_2O_3$/Cu as the function of the thickness of Cu. Black (red) points in (a) and (b) indicate the efficiency/coefficient for direct (inverse) effect. Error bars indicate the error propagation in the calculation.